\documentclass{ws-ijmpcs}
\begin{document}
\markboth{H. Velten \& J. Beltran Jimenez \& F. Piazza}
{Limits to anomalous speed of GW }

%
\catchline{}{}{}{}{}
%

\title{Limits on the anomalous speed of gravitational waves from binary pulsars}

\author{Hermano Velten}

\address{Universidade Federal do Espírito Santo, Av. Fernando Ferrari\\
Vit\'oria, Esp\'irito Santo, CEP 29075-910,
Brazil\\
velten@pq.cnpq.br}

\author{Jose Beltr\'an Jim\'enez and Federico Piazza}

\address{Aix Marseille Univ, Universit\'e de Toulon, CNRS, CPT, Marseille, France\\
jose.beltran@cpt.univ-mrs. \,\, fedosquare@gmail.com}

\maketitle

\begin{history}
\received{Day Month Year}
\revised{Day Month Year}
\published{Day Month Year}
\end{history}

\begin{abstract}
A large class of modified theories of gravity used as models for dark energy predict a propagation speed for gravitational waves which can differ from the speed of light. This difference of propagations speeds for photons and gravitons has an impact in the emission of gravitational waves by binary systems. Thus, we revisit the usual quadrupolar emission of binary system for an arbitrary propagation speed of gravitational waves and obtain the corresponding period decay formula. We then use timing data from the Hulse-Taylor binary pulsar and obtain that the speed of gravitational waves can only differ from the speed of light at the percentage level. This bound places tight constraints on dark energy models featuring an anomalous propagations speed for the gravitational waves.
\keywords{Gravitational waves; Modified Gravity; Binary pulsars.}
\end{abstract}

\ccode{PACS numbers:}

\section{Introduction}	

The direct detection of gravitational waves by the {\it Laser Interferometer Gravitational-Wave Observatory} (LIGO) has inaugurated the era of Gravitational Waves (GWs) astronomy \cite{Abbott:2016blz}. This breakthrough in the observation of the universe has taken place more than four decades after the discovery of the Hulse-Taylor binary pulsar \cite{Hulse1975}, which gave a first indirect evidence for the existence of GWs. The continuous timing measurements over the years revealed an orbital period decay in agreement with the emission of GWs as predicted by General Relativity (GR). In fact, the whole set of directly observed and inferred parameters in subsequently discovered binary pulsars have allowed to further confirm the validity of GR.

However, it is well known that cosmological observations  have revealed phenomena that call for new types of matter and/or modifications of GR as a theory of gravity. On the one hand, we need dark matter in order to explain, e.g., weak lensing and structures formation. Another major discovery in modern cosmology is the late-time accelerating expansion of the universe, what is broadly called \emph{dark energy}. In both cases, efforts have been made to explain the observations without additional components, but by modifying the gravitational interaction on the appropriate scales.  In particular, for cosmic acceleration a plethora of different infrared modifications of gravity have been extensively explored. A large class of these modified gravity theories give rise to non-trivial effects for the GWs propagating in cosmological scenarios, which can be divided into a modification of the GWs source (mainly the coupling constant to matter $G_{\rm gw}$) and an anomalous propagation speed $c_T\ne 1$. Our goal will be to use binary system observations to obtain bounds on these two effects and, thus, put constraints on dark energy models. It is very important to notice that, for a large class of theories, our bounds will apply even to models featuring screening mechanisms \`a la Vainshtein because the anomalous GW speed persists even inside screened environments\cite{Jimenez:2015bwa}.

\section{Modifying the radiative sector of gravitational theories}

As explained above, we will focus on theories of gravity displaying an anomalous propagation speed for GWs as well as a coupling constant different from the usual Newton's constant $G_N$ appearing in the corresponding Poisson equation (i.e., determining the orbits). Thus, we will consider the following effective lagrangian for GWs:
\begin{equation} \label{theory}
{\cal L} \ =  \frac{1}{64 \pi G_{\rm gw}}\sum_{\alpha=+,\times}\left[\frac{1}{c_T^2}\dot \gamma_{\alpha}^2  -  \vert\vec{\nabla} \gamma_{\alpha}\vert^2 \right]\, ,
\end{equation}
where $+, \times$ represent the two polarizations of the GWs. A few comments are in order here. We will assume that only the radiative sector differs from GR. Therefore it is assumed that the sector of the potential gravitons responsible for the gravitational bound of the system remains unaffected. Of course, in a modified gravity theory, such a sector is also expected to be modified, but Solar System experiments tightly constrain the corresponding PPN parameters so that it is safe (if not necessary) to assume that it is the same as in GR. On the other hand, we assume that $c_T$ is constant, direction- and polarization-independent for the scales of relevance in our problem. This is reasonable if the value of $c_T$ is determined by the cosmological evolution where only time translations are broken so that $c_T$ can only depend on time and vary on cosmological timescales, so that for the binary system scales it can be safely taken as a pure constant.


%


\section{Binary pulsar constraints}

The so called Damour-Deruelle post-Keplerian parameters \cite{DD} are written in terms of the Keplerian parameters and the unknown masses of the pulsar ($m_p$) and its companion ($m_c$). There are two relativistic effects which depend on the potential sector of the theory and therefore are not subjected to the modification of the radiative sector. Namely, the secular advance of periastron 
\begin{equation}
\dot{\omega} = 3 G_N^{2/3} c^{-2} (2\pi P_b)^{-5/3}(1-\epsilon^2)^{-1}(m_p+m_c)^{2/3} 
\end{equation}
and the amplitude of the Einstein delay
\begin{equation}
\gamma = G_N^{2/3} c^{-2} \epsilon (2\pi G_N)^{1/3} m_c (m_p+2m_c)(m_p+m_c)^{-4/3}.
\end{equation}
Here, the Keplerian parameters $\epsilon$ and $P_b$ are the eccentricity and period of the orbit, respectively. For details on obtaining the post-Keplerian parameters see Ref. \cite{Will:2014xja}.  

So far, it is clear that, from the pulsar timing observations, the measurement of $\dot{\omega}$ provides a way to constrain the total mass of the system $m_p + m_c$. Then, by obtaining the value of the Einstein delay $\gamma$ one is able the solve for the two masses. This is how binary pulsars became a powerful tool to test gravity theories and, in particular, to confirm the validity of GR predictions by using the orbital period decay $\dot{P_b}$ (see Ref. \cite{Wex:2014nva} for details). In fact, by using the masses values $m_p$ and $m_c$ obtained from $\dot{\omega}$ and $\gamma$ there is a remarkable agreement between the value of $\dot{P}_b$ predicted by GR and the measured one. 

Now we will show how modifications in the radiative sector will modify the expression for the orbital period decay $\dot{P_b}$. At this point we want to revisit, step by step, the standard derivation of the quadrupole emission formula. Firstly, we estimate the energy flux of a GW across a spherical surface at large distance $r$ from the source. The standard expression can be modified, essentially, 
by dimensional analysis. By rescaling the time as $\partial_{t}= c_T \partial_{t'}$, in such a way that the usual GR formulae can be applied straightforwardly, we find \cite{Maggiore:1900zz}
\begin{equation} \label{8}
\frac{d E}{d t}= \frac{r^2}{32 c_T \pi G_{\rm gw}}\int d\Omega \left\langle \partial_t \gamma_{ij} \partial_t \gamma_{ij}\right\rangle\, ,
\end{equation}
where $\langle \dots \rangle$ is the average over a spacetime region much larger than the GW wavelength. 
Also, at the lowest (quadrupole) order in the velocity expansion, the radiated amplitude of GWs from a given source is obtained with the usual known formula, but taking into account the replacement $G_N\rightarrow G_{\rm gw}$ and the different retarded time at which the source is evaluated, 
\begin{equation}
[\gamma_{ij}]_{quad} \ = \ \frac{2 G_{\rm gw}}{r} \ Q_{ij}^{TT} \!\left(t - \frac{r}{c_T}\right),
\end{equation}
where $Q_{ij}^{TT} $ is the transverse-traceless projection of the quadrupole moment $Q_{\ij}$ of the source. As in the standard calculation, the way such a projection is made depends on the direction of the GW and this should be taken into account when calculating the surface integral~\eqref{8}. The total power emitted reads therefore
\begin{equation}
P_{quad}=\frac{ G_{\rm gw}}{5 c_T} \left\langle  \dddot{Q}_{ij} \dddot{Q}_{ij}\right\rangle.
\label{Pquad}
\end{equation}

Equipped with the above modified quadrupole formula, we can straightforwardly obtain the expression for the decrease of the orbital period as
\begin{align} \label{Pdot}
\dot{P}_b =&-\left(\frac{G_{\rm gw}}{G_N}\frac{c}{c_T}\right) \frac{192 \pi G^{5/3}_N}{5 c^5}  \left(\frac{P_b}{2 \pi}\right)^{-\frac53}(1-e^2)^{-\frac72}\\ \nonumber
&\times\left(1+\frac{73 e^2}{24}+\frac{37 e^4}{96}\right) m_p m_c (m_p + m_c)^{-1/3},
\end{align}
where we have temporarily reintroduced the dimensional speed of light $c$.  Thus, we can now use binary pulsars observations  to constrain the combination $\frac{G_{\rm gw}}{G_N}\frac{c}{c_T}$.

Before obtaining the desired constraints, it is important to notice that the above equation is calculated in the orbiting system reference frame which is accelerated with respect to the solar system barycenter frame \cite{Damour1991}. This effect, known as Shklovskii effect, gives an extra $\Delta \dot{P}_{b,gal}=-0.027 \pm 0.005 \times 10^{-12}$ which should be subtracted in order to match the observed one.


We use the most accurate available data on $\dot P_b$, those of the Hulse-Taylor pulsar (PSR B1913+16), with the orbital parameters shown in Table \ref{Table1}~\cite{Weisberg2010}. 
In Fig. 1 we construct the usual mass-mass plot showing that all the observed parameters only coincide in a small region, determining that way the masses of the system and confirming the predictions of GR. The errors in the measurements are at the percentage level and, thus, only deviations of the same order with respect to GR are allowed. Therefore, the binary pulsar data will give us a constraint on $\frac{G_{\rm gw}}{G_N}\frac{c}{c_T}$ at the same level. More precisely, the $1-\sigma$ region of the observed parameters yields the following constraint for our modifications of the GWs
\begin{equation}
0.995\lesssim \frac{G_{\rm gw}}{G_N}\frac{c}{c_T}\lesssim 1.00
\end{equation}
which gives the advertised fact that only deviations from GR at the percentage are allowed. Moreover, if we further assume that $G_{\rm gw}$ is actually equal to $G_{\rm N}$, then we see that the speed of GWs can only differ from that of photons at the $10^{-2}$ level. 

\begin{figure}[t]
\includegraphics[width=0.95\textwidth]{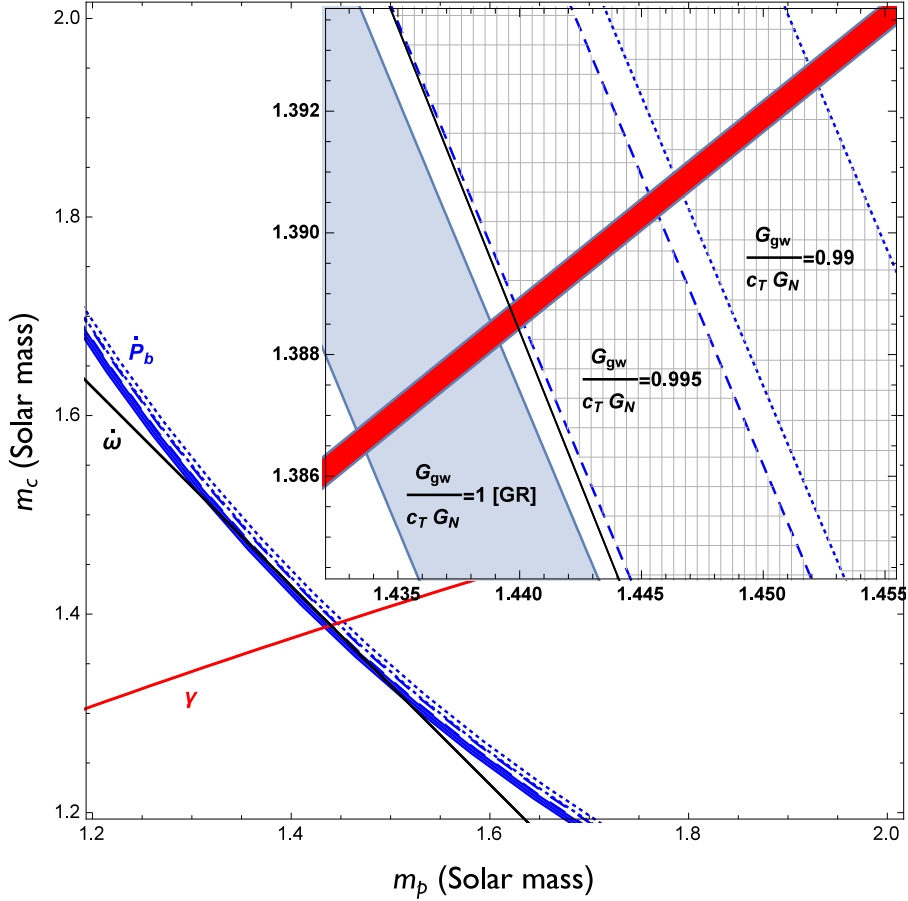}\\
\caption{Mass-mass diagram for PSR B1913+16 (the Hulse-Taylor pulsar) based on the post-Keplerian parameters $\dot{w}$ (black), $\gamma$ (red) and $\dot{P}_b$ (blue). Varying the combination $c_T G_{{\rm gw}}/G_N$ amounts to shifting the 1-$\sigma$ stripe of $\dot{P}_b$. }
\label{fig1}
\end{figure}



\begin{table}[h]
\centering
\begin{tabular}{ccc}
  \hline
  Parameter &Description & Value \\
 \hline
  $e$ & eccentricity & 0.6171334(5) \\
 $P_b({\rm days})$ & period & 0.322997448911(4) \\
  $\dot{w} ({\rm deg/ yr})$  & periastron advance& $4.226598(5)$ \\
  $\gamma ({\rm ms})$ & Einstein delay & $4.2992(8)$ \\
	$\dot{P}_b$ & period decay& $-2.423(1) \times 10^{-12}$ \\
	\hline
\end{tabular}
\caption{Orbital parameters for PSR B1913+16.} 
\label{Table1}
\end{table}

\section{Discussion}

We have obtained the constraints imposed by the Hulse-Pulsar binary system measurements on a modification of the radiative sector of gravity. In the case that the only modification is a change in the propagation speed of GWs, we have seen that such a speed can only differ from the speed of light at the  $10^{-2}$ level. The next natural step is to use direct detections of GWs to put constraints on the propagation speed of GWs. A first attempt was made in \cite{Blas:2016qmn} by simply looking at the time difference in the arrival of the GW to both LIGO detectors which gives the constraint $c_T\lesssim 1.7$. Another possibility will be to use the post-Newtonian phase of the GW signal, although the sensitivity of the detected signals do not allow to improve the constraint obtain here. However, the increased precision achieved with more interferometers will allow to obtain tighter constraints. Obviously, detecting an electromagnetic counterpart would also dramatically improve the constraints on $c_T$ and in \cite{Bettoni:2016mij} another method was proposed to constraint $c_T$ by simultaneously monitoring the electromagnetic and GW signals of binary systems with neutron stars.

\section*{Acknowledgments}
HV thanks support from CNPq and FAPES. This work has been funded by the 
A*MIDEX project (n ANR-11-IDEX-0001-02) funded by
the  "Investissements d'Avenir" French Government pro-
gram, managed by the French National Research Agency
(ANR). JBJ also acknowledges MINECO  (Spain) projects
FIS2014-52837-P and Consolider-Ingenio MULTIDARK
CSD2009-00064.

\end{document}